\documentclass{article}
\usepackage{spconf,amsmath,graphicx}

\usepackage{amsmath}
\usepackage{relsize}
\usepackage{amssymb}
\usepackage{amsfonts}
\usepackage{hyperref}
\usepackage{booktabs}
\usepackage{multirow}
\usepackage[noend,ruled,vlined]{algorithm2e}
\usepackage[noend]{algpseudocode}
\usepackage{subcaption}

\usepackage[normalem]{ulem}
\usepackage{etoolbox}
\usepackage{tikz}
\usetikzlibrary{tikzmark}
\usetikzlibrary{calc}

\errorcontextlines\maxdimen
\newcommand{\ALGtikzmarkcolor}{black}
\newcommand{\ALGtikzmarkextraindent}{4pt}
\newcommand{\ALGtikzmarkverticaloffsetstart}{-.5ex}
\newcommand{\ALGtikzmarkverticaloffsetend}{-.5ex}
\makeatletter
\newcounter{ALG@tikzmark@tempcnta}

\newcommand\ALG@tikzmark@start{%
    \global\let\ALG@tikzmark@last\ALG@tikzmark@starttext%
    \expandafter\edef\csname ALG@tikzmark@\theALG@nested\endcsname{\theALG@tikzmark@tempcnta}%
    \addtocounter{ALG@tikzmark@tempcnta}{1}%
}

\def\ALG@tikzmark@starttext{start}
\newcommand\ALG@tikzmark@end{%
    \ifx\ALG@tikzmark@last\ALG@tikzmark@starttext
    \else
        \tikzmark{ALG@tikzmark@end@\csname ALG@tikzmark@\theALG@nested\endcsname}%
        \tikz[overlay,remember picture] \draw[\ALGtikzmarkcolor] let \p{S}=($(pic cs:ALG@tikzmark@start@\csname ALG@tikzmark@\theALG@nested\endcsname)+(\ALGtikzmarkextraindent,\ALGtikzmarkverticaloffsetstart)$), \p{E}=($(pic cs:ALG@tikzmark@end@\csname ALG@tikzmark@\theALG@nested\endcsname)+(\ALGtikzmarkextraindent,\ALGtikzmarkverticaloffsetend)$) in (\x{S},\y{S})--(\x{S},\y{E});%
    \fi
    \gdef\ALG@tikzmark@last{end}%
}

\apptocmd{\ALG@beginblock}{\ALG@tikzmark@start}{}{\errmessage{failed to patch}}
\pretocmd{\ALG@endblock}{\ALG@tikzmark@end}{}{\errmessage{failed to patch}}
\makeatother

\RequirePackage[textsize=scriptsize]{todonotes}

\newcommand{\vek}[1]{{\mathbf {#1}}}
\newcommand{\vx}{{\vek{x}}}
\newcommand{\vy}{{\vek{y}}}
\newcommand{\vyhat}{{\hat{\vek{y}}}}
\newcommand{\vphat}{{\hat{\vek{p}}}}
\newcommand{\vp}{{\vek{p}}}

\newcommand{\vq}{{\vek{q}}}
\newcommand{\ve}{{\vek{e}}}

\newcommand{\score}{{\text{score}}}

\def\x{{\mathrm x}}
\def\y{{\mathrm y}}
\def\p{{\mathrm p}}
\def\q{{\mathrm q}}
\def\e{{\mathrm e}}

\newcommand{\asrModel}{{\mbox{$\mathcal{A}$}}}

\newcommand{\selectedSet}{{\mbox{$\mathcal{D}$}}}

\def\seed{{\cal S}}
\def\corpus{{\cal U}}

\def\testSet{{\cal T}}
\def\selectedSentences{{\cal Y}}
\def\budget{{\cal B}}
\def\errorModel{{\cal E}}
\def\phoneVocab{{\cal P}}

\def\hypotheses{{\cal H}}
\def\references{{\cal R}}

\title{Error-driven Fixed-Budget ASR Personalization for Accented Speakers} 

\name{Abhijeet Awasthi, Aman Kansal, Sunita Sarawagi, Preethi Jyothi}
\email{\{awasthi,amankansal,sunita,pjyothi\}@cse.iitb.ac.in}
\address{Department of Computer Science and Engineering, IIT Bombay, India}

\begin{document}
%
\maketitle
\begin{abstract}
We consider the task of personalizing ASR models while being constrained by a fixed budget on recording speaker specific utterances. Given a speaker and an ASR model, we propose a method of identifying sentences for which the speaker's utterances are likely to be harder for the given ASR model to recognize. We assume a tiny amount of speaker-specific data to learn phoneme-level error models which help us select such sentences. We show that speaker's utterances on the sentences selected using our error model indeed have larger error rates when compared to speaker's utterances on randomly selected sentences. We find that fine-tuning the ASR model on the sentence utterances selected with the help of error models yield higher WER improvements in comparison to fine-tuning on an equal number of randomly selected sentence utterances. Thus, our method provides an efficient way of collecting speaker utterances under budget constraints for personalizing ASR models. Code for our experiments is publicly available~\cite{awasthia14:online}.

\end{abstract}
\section{Introduction}
\label{sec:intro}
Even as state of the art ASR models provide impressive accuracy on mainstream speakers, their accuracy on accented speakers is often drastically low. On a state of the art ASR system, we observed WERs ranging from 11.0 to 53.0 across speaker accents from eight different Indian states in contrast to a WER of less than 4.0 on native English speakers. 
With the proliferation of voice-based interfaces in several critical mobile applications, it is imperative to provide quick and easy personalization to individual users for fairness and universal accessibility.  Recent work~\cite{Shor_2019,sim2019investigation} has established that fine-tuning with speaker-specific utterances is an effective strategy for personalizing an ASR model for accented speakers.  In this paper, we address the complementary question of how to efficiently collect such speaker utterances.   We reduce speaker effort significantly by carefully selecting the set of sentences for recording speaker utterance data for fine-tuning.

Existing work on selecting sentences is surprisingly limited to only strategies like enforcing phonetic or word diversity among a selected set of sentences~\cite{mendoncca2014method,wu2007data,wei2014submodular}.  In contrast,  the reverse problem of selecting utterances to transcribe from an existing unlabeled utterance corpus is called the active learning problem and has been extensively studied~\cite{riccardi2005active,yu2010active,hamanaka2010speech,nallasamy2012active,fraga2015active,syed2017active}.
Our problem is better motivated to the task of personalizing to diverse user accents where large unlabeled utterances are non-existent, and labeled data has to be collected by recording utterances on selected sentences.

We present a new algorithm for selecting a fixed number of  sentences from a large text corpus  for fine-tuning an existing ASR model to a specific speaker accent.  We show that it is important to select sentences on which the current model is likely to be erroneous.   When selecting a batch of sentences, we also need to ensure that the selected batch is phonetically diverse and representative of the test distribution.  We therefore select sentences by a combination of two terms:  the phoneme-level predicted error probability, and a second term for maintaining balance  across the selected phonemes.  
We show that our phoneme-level error detector trained on a small random seed labeled set is surprisingly effective in filtering out sentences whose utterance incur high WER.   For example, the utterance of the top-100 sentences filtered by using our error model incur even upto 100\% higher WER than utterances of randomly selected sentences (Table~\ref{tab:wers}).

Our work takes us a step closer towards inclusivity on the recent advances in ASR technology.
Even within the broad category of Indian accents, we observe WERs on accented English ranging from 11.1 of a mainstream Hindi male speaker to 27.1 of a far-east Assamese female. After fine-tuning each with just 135 and 150 of our selected sentences we reduce their WERs to 8.2 and 19.0 respectively, whereas the corresponding random selection would require 250 and 180 sentences for the same drop.

\section{Related Work}

Active learning for speech recognition aims at identifying the most informative utterances to be manually transcribed from a large pool of unlabeled speech. This topic has been extensively explored on a number of different fronts, including the use of uncertainty-based sampling to select informative speech samples~\cite{riccardi2005active,yu2010active,hamanaka2010speech,nallasamy2012active}, active learning for low-resource speech recognition~\cite{fraga2015active,syed2017active}, combined active and semi-supervised learning~\cite{drugman2019active} and active learning for end-to-end ASR systems~\cite{yuan2019gradient,malhotra2019active}. 

In active learning, the goal is to select informative speech samples that are subsequently transcribed, while our work focuses on the reverse problem of selecting informative sentences that are subsequently recorded as speech. The latter lends itself well to building personalized ASR models where a small number of speaker-specific speech samples are used to personalize speaker-independent ASR models. Existing work on data selection is limited to selecting text based on phonetic diversity or word-diversity~\cite{wu2007data,wei2014submodular,mendoncca2014method}.  In contrast, our selection method is adaptive to the observed accent errors of a pretrained ASR model, and we show that it provides much greater gains.
Many recent work investigate methods of accent adaptation~\cite{Shor_2019,kat1999fast,sun2018domain,jain2018improved} but they all assume availability of labeled data in the target accent.  Ours is the only work we are aware of that focuses on the problem of efficiently collecting such labelled data for accented speakers. 

\section{Our Method}
We assume a pretrained ASR model $\asrModel$, a tiny seed set $\seed = \{(\vx_1,\vy_1), (\vx_2,\vy_2), \ldots (\vx_S,\vy_S)\}$ comprising of pairs of speaker's utterance ($\vx_i$) and reference transcript ($\vy_i$), a large corpus of sentences $\corpus = \{\vy_1, \vy_2, \ldots \vy_U\}$.
For fine-tuning $\asrModel$, we wish to record a few additional speaker's utterances on a subset of sentences in $\corpus$. We assume a budget $\budget$ on number of sentences for which utterances can be recorded.  Let the collected utterances and the corresponding sentences be represented by $\selectedSet = \{(\vx_1,\vy_1), (\vx_2,\vy_2), \ldots (\vx_\budget,\vy_\budget)\}$. 
Our goal is to find a way for selecting the sentences $\{\vy_1,\vy_2,\ldots,\vy_\budget\} \subset \corpus$ such that fine-tuning $\asrModel$ on $\seed \cup \selectedSet$ yields better test performance 
in comparison to random selection.  

A simple baseline is to select sentences uniformly at random from $\corpus$.  Intuitively, we hope to perform better than this baseline by selecting $\selectedSet$ where $\asrModel$ is likely to be erroneous.
We next present the design of our method  that makes this judgement based on the seed set $\seed$. 
Our key idea is to learn an error detector $\errorModel$ that helps us spot phonemes in a given sentence which are likely to get misrecognised when the sentence's utterance is fed as an input to our pre-trained ASR model $\asrModel$.   In Section~\ref{sec:error} we present the design of our error detection module.  Then in Section~\ref{sec:alg} we present the algorithm that performs the final set selection.

\subsection{The Error Detection Model} 
\label{sec:error}
We first convert a sentence to a sequence of phonemes using a pretrained grapheme-to-phoneme convertor \footnote{\url{https://github.com/Kyubyong/g2p}}.  Let $\vp_i = [p_i^1,p_i^2,\ldots,p_i^{n_i}]$ be the phonemes of sentence $\vy_i$, where $p_i^j \in \phoneVocab$, the phoneme alphabet and $n_i$ be the number of phonemes in sentence $\vy_i$.  
Let $\ve_i = [\e_i^1, \e_i^2, \ldots, \e_i^{n_i}]$ represent a sequence of same length as $\vp_i$ such that $\e_i^j = 1$ if phoneme $p_i^j$ gets misrecognized by the pretrained ASR model $\asrModel$. The error model $\errorModel$ outputs the error probabilities $\vq_i = [\q_i^1,\q_i^2,\ldots,\q_i^{n_i}]$, where ${\q_i^j = \Pr(\e_i^j=1 | \vp_i)}$. 

\begin{figure}
        \centering
        \includegraphics[scale=0.30]{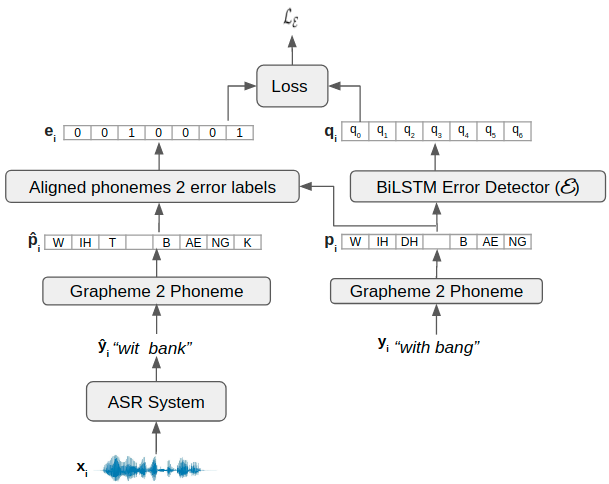}
        \caption[]
        {Training error model $\errorModel$ where $(\vx_i,\vy_i) \in \seed$, $\hat{\vy_i}$ is recognized by $\asrModel$.  Their corresponding phonemes are aligned to assign an error label to each phoneme of $\vy_i$ which then serves as labeled data for the error detection model } 
        \label{fig:train_error_model}
\end{figure}

\paragraph*{\bf Training the Error Model:}
We use the seed set $\seed$ to train $\errorModel$.
Figure~\ref{fig:train_error_model} presents an overview of this training.
 First, we invoke the pretrained ASR model $\asrModel$ on the utterances in $\seed$ to obtain a set of hypotheses $\hypotheses = [\vyhat_1, \vyhat_2, \ldots, \vyhat_S]$. Let the corresponding references be $\references = [\vy_1, \vy_2, \ldots, \vy_S]$. 
 Using a phone-aware edit distance algorithm as in \cite{power-asr}, we align the phoneme representation $\vphat_i$ of each hypothesis $\vyhat_i$ with the phoneme representation $\vp_i$ of the corresponding reference $\vy_i$. Using these alignments, we obtain the error sequence $\ve_i$ such that $\e_i^j=0$ if the token aligned with $\vp_i$ at position $j$ is the same as $p_i^j$. Otherwise, $\e_i^j=1$, representing that the phone $p_i^j$ at position $j$ in the reference $\vp_i$ got misrecognised in the hypothesis $\vphat_i$. The example in Figure~\ref{fig:train_error_model} clearly explains these steps. 
The training of the error model can now be posed as a sequence labeling problem. At each position $j$ in the reference phone sequence $\vp_i$ we minimize the cross-entropy loss on the binary label $\e_i^j$. The training loss is thus:\begin{equation}
    \label{equation:errorloss}
    \mathcal{L_{\errorModel}} = \sum_{i\in \seed} \sum_{j=1}^{n_i} \log(\Pr(\e^j_i=1|\vp_i))
\end{equation}  

\paragraph*{Model Architecture} We implemented the error model $\errorModel$ as a 4-layer bi-LSTM  which takes as input feature representation of phonemes, followed by a ReLU activation, a linear layer and a softmax activation to produce error probabilities $\q_i^j$. The hidden states of the bi-LSTM are of size 64. Each phoneme is represented by a concatenation of three types of learned embeddings: 64-dimensions for each phoneme, 8-dimensions for Vowel/consonant feature of the phoneme,  and 8-dimensions for its phoneme-type grouped as {monophthongs, diphthongs, stops, afficates, fricatives, nasals, liquids, semivowels}.
We train
using Adam optimizer for 100 epochs with a learning rate of 3e-4, batch size of 1 and early stopping using a small dev set. 

\subsection{Sentence selection algorithm}
\label{sec:alg}
We select the set $\selectedSentences$ of $\budget$ sentences from  the corpus $\corpus$ iteratively.  Having selected the first  $i$ sentences in $\selectedSentences$, we score the remaining sentences in $\corpus$ and choose the highest scoring of those.  If sentences are scored purely based on their predicted error probability, we might overly bias the selected set to errors observed in the small seed set.  We therefore introduce a second term to penalize over-represented phonemes. \\
Let  $n_\pi(\selectedSentences)$ denote the number of occurrences of a phoneme $\pi$ in set $\selectedSentences$.  Each phoneme $\pi$ in the next $(i+1)$-th sentence to be selected is assigned a weight $c_\pi$ based on how popular $\pi$ already is in $\selectedSentences$ --- a phoneme already well-represented gets a small weight than an under-represented phoneme.  In other words, we wish to enforce a diminishing return effect that characterizes sub-modular set-scoring function.  For such functions, incremental selection algorithms like ours are known to provide competitive approximation guarantees~\cite{shinohara2014submodular}.  Accordingly, we define $c_\pi(\selectedSentences, \vy) = f(n_\pi(\selectedSentences\cup \vy)) - f(n_\pi(\selectedSentences))$ where $f$ is a sub modular function.  We choose $f(n)=1 - \exp(-\frac{n}{\tau})$ where $\tau$ is a hyper-parameter. We set $\tau=500$ for all our experiments.  
We combine the above $c_\pi$ term with the predicted error probabilities to define a score for each candidate sentence $\vy$ to be included in an existing set $\selectedSentences$ as per Equation~\ref{eq:score}. Here $\vp$ denotes the corresponding phoneme sequence of $\vy$ and $n$ is number of phonemes in $\vp$.
\begin{align}
\label{eq:score}
     \score(\vy,\vp,\selectedSentences) = \frac{1}{n}\; \mathlarger{\sum}_{\pi\in \phoneVocab}  c_\pi(\selectedSentences,\vy) \sum_{j:p_j = \pi} P_\errorModel(e^j=1|\vp)
\end{align}
For each phoneme in the sentence, the first term $c_\pi$ measures the diminishing gains due to including the phonemes in $\vy$ to existing counts in $\selectedSentences$, while the second term 
encourages higher scores if $\asrModel$ is likely to misrecognize different occurrences of that phoneme when uttered by an accented speaker. The scores are normalized by  
the phoneme sequence length to reduce bias towards longer sentences. Our overall algorithm appears in Algorithm~\ref{algo1:sample_trainset}.

\begin{algorithm}
\begin{small}
\textbf{Require} $\corpus$, $\budget, \seed, \asrModel$

$\errorModel \gets$ Train error model using \seed,\asrModel\ (Section~\ref{sec:error}) 

$\selectedSentences \gets \Call{EmptySet}{}$

\For{$i \gets 1$ to $\budget$}{
    \For{$\vy \in \corpus - \selectedSentences $}{

        $\vp \gets \Call{Grapheme2Phoneme}{\vy}$
        
        Calculate $\score(\vy,\vp,\selectedSentences)$ using Equation~\ref{eq:score}.
        
%
}
    Add the highest scoring $\vy$ from above to $\selectedSentences$.
    
}
$D \gets $ Collect speaker utterances $\vx_i$ for each $\vy_i \in \selectedSentences$

Fine-tune ASR model \asrModel\ on $D \cup \seed$.

\caption{Personalizing the ASR model \asrModel}
\label{algo1:sample_trainset}
\end{small}
\end{algorithm}

\algnewcommand{\LineComment}[1]{\State \(\triangleright\) #1}
\renewcommand{\algorithmicrequire}{\textbf{Input:}}
\renewcommand{\algorithmicensure}{\textbf{Output:}}

\section{Experiments}
We experiment with fifteen speakers spanning different regions and gender on a pre-trained English ASR model. Code for our experiments is publicly available~\cite{awasthia14:online}.

\paragraph*{The ASR model}
We use QuartzNet-15x5 \cite{kriman2020quartznet} as our pre-trained ASR model $\asrModel$, that was trained on LibriSpeech \cite{panayotov2015librispeech} for 400 epochs using CTC-loss \cite{graves2006connectionist} and has a greedy WER of 3.90 on test-clean of LibriSpeech. The QuartzNet-15x5 architecture is a more compact variant (19M params) of the widely used JasperDR-10x5 architecture~\cite{li2019jasper} (333M params), which is fully convolutional with residual connections. The pre-trained model is finetuned on speaker-specific utterances by minimizing CTC-loss using the NovoGrad optimizer \cite{novograd} for 100 epochs with a linearly decaying learning rate of $10^{-5}$, batch size of 16, and early stopping based on a dev set. 

\setlength\tabcolsep{1.5pt}

\begin{table}
\centering
\scalebox{0.95}{
\begin{tabular}{|l|r|r|r|r|r|r|r|}
\hline
 & \multicolumn{7}{|c|}{IndicTTS} \\ \hline
 &    & \multicolumn{3}{|c|}{$\budget=$100} &  \multicolumn{3}{|c|}{$\budget=$500} \\ \hline
Spkr & Pre-trained & Rand & Diverse & Our & Rand & Diverse & Our \\ \hline
Kn, M & 18.7 & 13.5 & 14.6 & \textbf{12.7} & 11.2 & 11.7 & \textbf{10.7} \\
Ml, M & 19.5 & 15.2  & 15.1 & \textbf{14.8} & 12.7 & 13.7 & \textbf{12.2} \\
Ra, M & 21.9 & 14.9 & 15.9 & \textbf{14.8} & 13.5 & 14.0 & \textbf{13.1} \\
Hi, M & 11.1 & 8.9 & 8.9 & \textbf{8.2} & 7.9 & 8.0 & \textbf{7.4} \\
Ta, M & 12.5 & \textbf{11.5} & 11.8 & \textbf{11.5} & 10.5 & 10.9 & \textbf{10.2}\\
As, F & 27.1 & 19.2 & 19.3 & \textbf{19.0} & 17.1 & 16.8 & \textbf{16.2} \\
Gu, F & 13.7 & 9.4 & 9.6 & \textbf{9.2} & 8.1  & 8.4 & \textbf{7.7} \\
Ma, F & 53.1 & 42.4 & 42.5 & \textbf{42.0} & 38.9 & 39.2 & \textbf{37.8}\\\midrule
 & \multicolumn{7}{|c|}{L2-Arctic} \\ \hline
 &    & \multicolumn{3}{|c|}{$\budget=$50} &  \multicolumn{3}{|c|}{$\budget=$100} \\ \hline
Spkr & Pre-trained & Rand & Diverse & Our & Rand & Diverse & Our \\ \hline
TLV & 44.8 & \textbf{37.7}    & 38.5  & 37.8  &             36.7 & 37.0 & \textbf{36.1} \\    
LXC & 37.1 &  \textbf{30.2}  &  30.8 & 31.0  &             29.5 & 30.1 & \textbf{29.1} \\ 
ERMS & 24.0 & \textbf{20.9}  & 21.2  & \textbf{20.9}  &              20.3 & 20.3 & \textbf{20.0} \\ 
HKK & 26.1 &  21.4  &  22.2 & \textbf{21.2}  &             \textbf{20.9} & 22.0 & 21.0  \\   
ABA & 24.5 &  \textbf{22.1}  & 22.4  & \textbf{22.1}  &             21.8 & 22.5 & \textbf{20.5}  \\   
\hline
\end{tabular}}
\caption{Test WER for each accent on the ASR model  finetuned with sets selected using three methods.  The first column is the speaker's regions, gender pair, the  second column is WER on pre-trained model, and remaining columns provide fine-tuned model's WER. All the numbers were averaged over atleast 3 random seeds.}
\label{tab:overall}
\end{table}

\paragraph*{Datasets}
We experiment on two public datasets: L2-Arctic \cite{zhao2018l2arctic} and IndicTTS \cite{indicTTS}. L2-Arctic borrows sentences from the CMU-Arctic dataset \cite{cmuarctic} but records non-native speakers of English. We consider speakers with Vietnamese (TLV), Mandarin (LXC), Spanish (ERMS), Korean (HKK) or Arabic (ABA) as their native language. The IndicTTS dataset comprises  
English utterances of speakers with diverse accents . We consider speakers with Kannada (Kn), Malayalam (Ml), Rajasthani (Ra), Assamese (As), Gujarati (Gu), Hindi (Hi), Manipuri (Ma) or Tamil (Ta) as their native language across both genders as shown in Table~\ref{tab:overall}. For each speaker the dataset is divided into 4 parts: the seed set $\seed$, the dev set used for early-stopping while finetuning the ASR model, the corpus $\corpus$ and the test set $\testSet$. 
The dev and seed set is set to 50 sentences each. For IndicTTS, the average size of the corpus $\corpus$ and the test set $\testSet$ is 4.3K and 1.9K sentences respectively. 
For L2-Arctic, the corpus $\corpus$ contains 690 sentences, while the test set $\testSet$ contains 340 sentences for all the speakers.
For training the error model $\errorModel$, we merge the dev set along with the seed set and keep aside 35\% of data which now serves as the dev set for the error model. The remaining 65\% of the data is used to train the error model.

\paragraph*{Comparison with Existing methods}
We compare our method with two other baselines: {\bf Random selection} of sentences  (``Rand") and {\bf Phonetically rich} sentences (``Diverse") selected using \cite{mendoncca2014method}'s method from the corpus $\corpus$. In order to ensure that different methods get the same total time budget, all the methods are given the same total duration as the $\budget$ sentences selected by random sampling.  The seed set $\seed$ is included for all the methods during finetuning.

\begin{figure}
        \centering
        \includegraphics[width=0.90\linewidth]{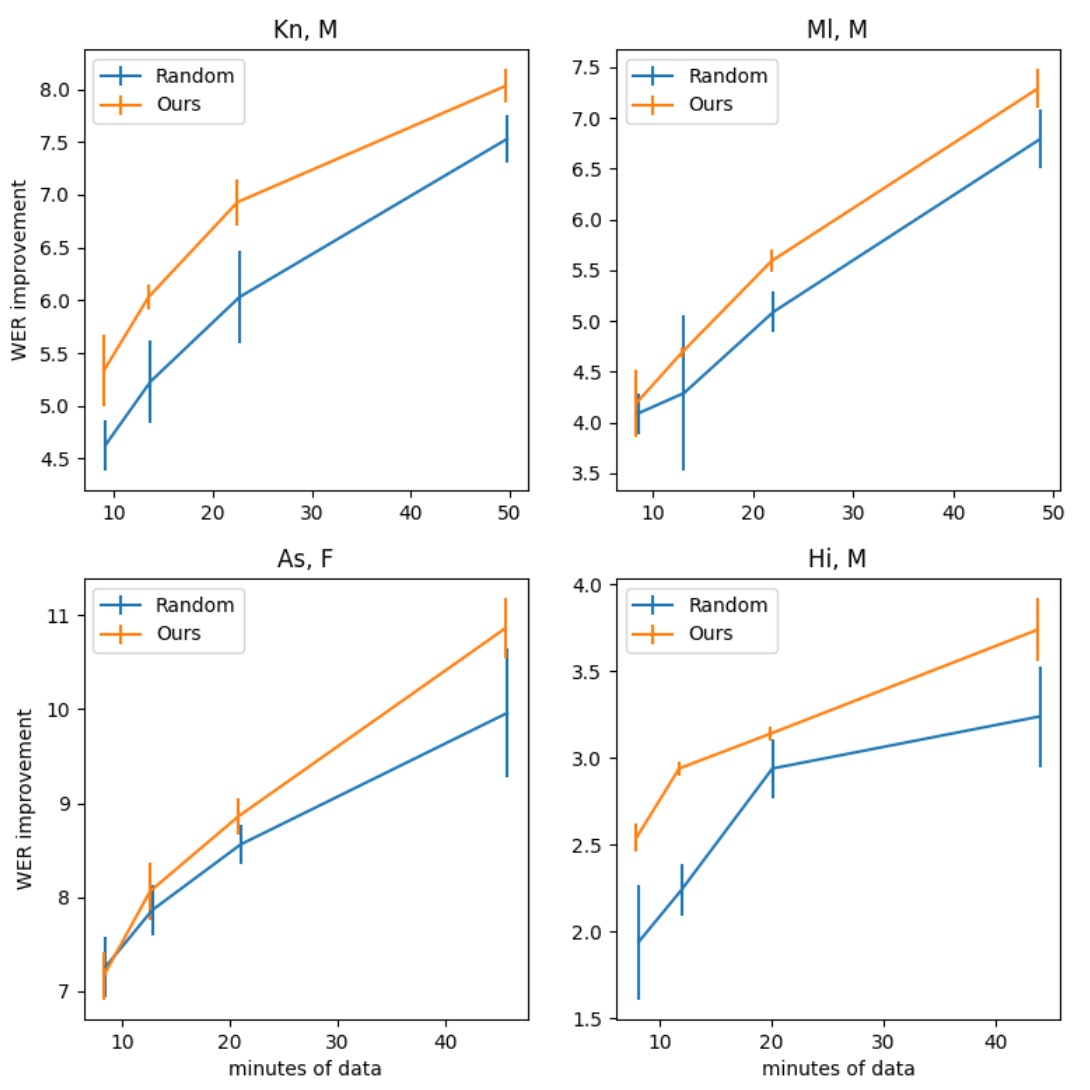}
        \caption[ Efficiency Figures ]
        { WER improvement (y-axis) vs Minutes of data used including the seed set (x-saxis)} 
        \label{fig:efficiency_plot}
\end{figure}

We present our comparisons in Table~\ref{tab:overall} that shows the WER of the fine-tuned model with a $\budget$ of 100 and 500 instances for IndicTTS, and 50 and 100 for the smaller L2-Arctic corpus. 
First observe that compared to the pre-trained model, fine-tuning even with 100 examples helps significantly. The WER reduces further as we increase the fine-tuning budget.  For the same time budget if sentences are selected using our method, the gains are significantly higher than random.  The  phonetic diversity baseline does not improve beyond random. We note here that our improvements over random on L2-Arctic are smaller compared to IndicTTS due to the smaller size of the selection corpus $\corpus$. \\ 
We present another perspective on the reduction of speaker effort that our method achieves in Figure~\ref{fig:efficiency_plot}. The y-axis is the WER improvement over the pre-trained model that finetuning achieves with varying amount of labeled data in minutes in the x-axis, using sentences selected by our method (orange) and random (blue). E.g. in the top-left figure, we see that we would require 14 minutes of data to achieve a WER reduction of 6.0 whereas random would require 22 minutes of data.

\begin{table}[]
\centering
\begin{tabular}{|l|r|r|r|}
\hline
Speaker & Random & Error Model & Skyline \\ \hline
Kn,M    & 14.8   & 21.5        & 58.6    \\ \hline
Ml,M    & 16.2   & 24.0        & 66.4    \\ \hline
Ra,M    & 17.1   & 26.0        & 67.5    \\ \hline
Hi,M    & 7.1    & 18.1        & 40.6    \\ \hline
Ta,M    & 8.4    & 16.5        & 52.7    \\ \hline
As,F    & 20.8   & 31.7        & 85.8    \\ \hline
Gu,F    & 10.3   & 19.4        & 58.1    \\ \hline
Ma,F    & 40.1   & 60.6        & 166.0   \\ \hline
\end{tabular}
\caption{WER of the pretrained ASR model on 100 sentences selected randomly or through the error model. Skyline represents selecting top-100 sentences ranked in the decreasing order of WER}
\label{tab:wers}
\end{table}

A primary reason our method performs better is that the error model $\errorModel$ enables us to prefer sentences where the pre-trained model is worse.
In Table~\ref{tab:wers}, we show WER of the pre-trained model on the top-100 sentences selected by our method.  We contrast their WER with randomly selected sentences.  Also, we create an (unachievable) skyline by selecting the WER of the top-100 highest error sentences based on actual predictions from the ASR model.  We see  that WER of the ASR model is higher on sentences selected by our method in comparison to random. 
Thus, the error models help us select sentences whose utterances are likely to be challenging for the ASR model. Finetuning on such challenging utterances allows better personalization of the ASR model.

\section{Conclusion and Future Work}
In this paper we presented a method of selecting sentences within a fixed budget that yield better personalization of existing ASR models to accented speakers than random selection.  Our key idea is to train a phoneme-level error detector using an initial seed set of the speaker's samples, and use that to further bias the selection towards sentences that manifest ASR errors. In future we would like to try our method to provide efficient personalization for dysarthric speech. \\

\noindent
{\bf Acknowledgement} We thank the team of the IndicTTS project at IIT Madras for providing us the datasets containing speech from a wide variety Indian accents. We also thank NVIDIA and L2-Arctic project for open sourcing their code and datasets.   This research was partly sponsored by IBM AI Horizon Networks - IIT Bombay initiative.  The first author was supported by Google PhD Fellowship.






\bibliographystyle{IEEEbib}
\bibliography{references}

\end{document}